\documentclass[aps,prd,twocolumn,notitlepage,nofootinbib,nobibnotes,superscriptaddress,amsmath,amssymb,preprintnumbers]{revtex4-1}
\usepackage{epsfig}
\usepackage{bm}
\usepackage{amssymb}
\usepackage{amsmath}
\usepackage[hidelinks]{hyperref}
\usepackage{xcolor}
\usepackage{subfigure}
\usepackage{float}
\usepackage{hyperref}
\graphicspath{{./Figures/}}

\definecolor{DeepGreen}{HTML}{2A7F62}
\definecolor{DeepBlue}{HTML}{00416A}
\definecolor{DeepOrange}{HTML}{A68C6C}
\definecolor{DeepRed}{HTML}{E63946}

\newcommand{\trento}{T\raisebox{-.5ex}{R}ENTo}

\begin{document}
\title{Mapping Nuclear Deformation with Differential Radial Flow in Heavy-Ion Collisions}

\author{Jie Zhu}
\affiliation{Institute of Particle Physics and Key Laboratory of Quark and Lepton Physics (MOE), Central China Normal University, Wuhan, 430079, China}
\affiliation{Fakult\"at f\"ur Physik, Universit\"at Bielefeld, D-33615 Bielefeld, Germany}

\author{Xiang-Yu Wu}
\email{xiangyu.wu2@mail.mcgill.ca}
\affiliation{Department of Physics, McGill University, Montreal, Quebec, H3A 2T8, Canada}

\author{Guang-You Qin}
\email{guangyou.qin@ccnu.edu.cn}
\affiliation{Institute of Particle Physics and Key Laboratory of Quark and Lepton Physics (MOE), Central China Normal University, Wuhan, 430079, China}

\begin{abstract}

In relativistic heavy-ion collisions, the radial flow of the fireball, usually characterized by transverse momentum spectra of final-state particles, encodes essential information about the hot and dense nuclear matter created in the collisions. However, the response of radial flow, including its $p_T$-differential structure $v_0(p_T)$ and longitudinal fluctuations $v_0(\eta)$, to intrinsic nuclear deformation remains unexplored. 
Using realistic $(3+1)$-dimensional viscous hydrodynamic calculations with \trento-3D initial conditions, we investigate how nuclear deformation affects the differential radial flow. 
We observe a clear, positive correlation between quadrupole deformation $\beta_2$  and radial flow: both magnitudes of $v_0$ and $v_0(p_T)$ are enhanced in central collisions when $\beta_2$ is increased. 
In contrast, the Pearson coefficient $\rho(n(p_T), [p_T])$ exhibits a universal step-like behavior across all collision systems and centralities.
Further analysis of longitudinal decorrelation of radial flow reveals a rich structure: in central collisions, large $\beta_2$ tends to suppress the decorrelation, whereas hexadecapole deformation $\beta_4$ tends to enhance it. Such decorrelation effect increases toward peripheral collisions. 
Our results demonstrate that precise measurements of radial flow, spanning transverse momentum and longitudinal dependences, can provide powerful, complementary constraints on nuclear deformation in high-energy nucleus-nucleus collisions.

\end{abstract}

\maketitle

\section{Introduction}\label{sec1}
Nuclei distant from magic numbers usually exhibit various types of deformation~\cite{Heyde:2016sop}, which are traditionally explored by comparing spectroscopic data with theoretical model calculations~\cite{Heyde:2011pgw, Bender:2003jk, Cline:1986ik,Yang:2022wbl}.
To quantify the geometric structure of nuclei, one commonly employs the Woods–Saxon distribution to characterize their shape and radial density profiles:
\begin{equation}
    \rho(r, \theta, \phi) \propto \frac{1}{1+\exp({r-R(\theta, \phi) \over a})}.
    \label{eq:defws1}
\end{equation}
Here $a$ represents the surface diffuseness parameter, and $R(\theta, \phi)$ denotes the nuclear radius which is usually expanded in spherical harmonics $Y_{l, m}$ as  
\begin{equation}
\begin{aligned}
    &R(\theta, \phi)
    =R_0 [1+\Sigma_{l, m} \beta_l \alpha_{l,m} Y_{l, m}(\theta, \phi)]
    \\
    &=R_0(1+\beta_2 [\cos\gamma Y_{2,0} + \sin\gamma Y_{2,2} ]+\beta_3 Y_{3,0}+\beta_4 Y_{4,0}\cdots),
\end{aligned}
    \label{eq:defws2}
\end{equation}
where $\beta_2$, $\beta_3$, and $\beta_4$ quantify the quadrupole, octupole and hexadecapole deformations, $\gamma$ is the triaxiality parameter that controls the relative ordering of the three radii~\cite{Jia:2021tzt}.
Complementary to traditional low-energy nuclear physics experiments, high-energy heavy-ion collision experiments at RHIC and the LHC offer a novel avenue to investigate the intrinsic structure of the nuclei~\cite{Jia:2022ozr, Jia:2021qyu,Ma:2022dbh,Jia:2022qgl, ChunJian:2024vdk,Giacalone:2025vxa, Luzum:2023gwy,Shou:2014eya}. In these collisions, a strongly coupled quark-gluon plasma (QGP), consisting of deconfined quarks and gluons, is created. This hot and dense medium exhibits strong collective flow, which efficiently translates the initial spatial geometry and its fluctuations in the colliding overlap region into the momentum anisotropy and corresponding fluctuations of the final-state hadrons. Leveraging the success of relativistic viscous hydrodynamics in describing collective flow phenomena, it becomes possible to constrain the underlying deformation parameters of the colliding nuclei, particularly in the regime of ultra-central collisions.

Indeed, extensive studies within the heavy-ion collision community have investigated deformed nuclei through the analysis of anisotropic flow~\cite{Wang:2024vjf,STAR:2024wgy,Li:2026igf,Xu:2025cgx,Mantysaari:2024uwn}, mean transverse momentum fluctuations~\cite{Jia:2021qyu,STAR:2024wgy,Zhang:2025hvi}, and their correlations~\cite{Wang:2024vjf, STAR:2024wgy, Jia:2021qyu, Jia:2021wbq,Fortier:2023xxy,Fortier:2024yxs,Zhao:2024lpc,Liu:2025fnq}. For instance, the STAR Collaboration~\cite{STAR:2024wgy} analyzed the ratios of the fluctuating observable $\langle v_2^2 \delta{p_T}\rangle$ in U+U versus Au+Au collisions and successfully extracted the deformation parameters of $^{238}\text{U}$ as $\beta_2^U = 0.297 \pm 0.015$ and $\gamma_U = 8.5^{\circ} \pm 4.8^{\circ}$, which are consistent with the values derived from rotational spectra~\cite{Pritychenko:2013gwa}. More recently, this approach has been extended to light nuclei, offering a novel means to probe exotic intrinsic geometries such as $\alpha$-cluster structures~\cite{Giacalone:2024luz,Ding:2023ibq,ATLAS:2025nnt,Giacalone:2024ixe,Hu:2025eid,Ke:2025tyv,Nijs:2025qxm,YuanyuanWang:2024sgp,Zhang:2024vkh}.

However, nuclear structure dictates not only the geometry but also the initial size and density gradients of the fireball, which drive the isotropic expansion dynamics of the QGP, manifesting as radial flow. Conventionally, radial flow is extracted by fitting final-state hadron $p_T$ spectra with a Blast-Wave model~\cite{ALICE:2019hno,STAR:2017sal}. However, this approach yields only global average parameters and is also sensitive to non-flow effects induced by jets and resonance decays~\cite{Saha:2025nyu}. 
To resolve these issues, a transverse momentum dependent radial flow observable, $v_0(p_T)$, was recently proposed~\cite{Schenke:2020uqq} and subsequently measured by the ATLAS~\cite{ATLAS:2025ztg} and ALICE~\cite{ALICE:2025iud} collaborations. 
Recent theoretical studies using the Blast-Wave model indicate that the mass ordering of $v_0(p_T)$ at low $p_T$ arises from intrinsic spectral shape fluctuations rather than azimuthal flow fluctuations, whereas the baryon-meson splitting at intermediate $p_T$ serves as a signature of the quark coalescence mechanism~\cite{Wan:2025rzg}. In contrast to anisotropic flow, which are sensitive to shear viscosity, $v_0(p_T)$ exhibits sensitivity to the bulk viscosity and the equation of state (EoS)~\cite{Liu:2025fbu}. Furthermore, a systematic study~\cite{Du:2025dpu} of $v_0(p_T)$ and its scaled counterpart showed that the initial effective nucleon width and off-equilibrium corrections ($\delta f$) also influence radial flow, particularly at LHC energies, while the observable remains insensitive to the late-stage hadronic rescattering. 

Although $v_0(p_T)$ is known to capture fluctuations related to isotropic expansion, its sensitivity to the initial nuclear geometry remains unexplored. In this work, we aim to investigate whether this differential radial flow observable can be utilized to constrain nuclear deformation parameters. 

The paper is organized as follows. 
In Sec.~\ref{sec2}, we briefly introduce the (3+1)-dimensional CLVisc viscous hydrodynamics framework, which is utilized to simulate event-by-event heavy-ion collisions at RHIC.
The numerical results are presented in Sec.~\ref{sec3}. 
We first demonstrate the calibration of model against experimental measurements, such as $dN/d\eta$, $p_T$ spectra and the mean transverse momentum $\langle [p_T] \rangle$ as a function of centrality.
We then systematically study the impact of nuclear structure on the radial flow observalbes, specifically $v_0$, $v_0(p_T)$ and the scaled radial flow $v_0(p_T)/v_0$ , across various collision systems and centrality bins. Furthermore, we analyze the Pearson correlation coefficient $\rho(n(p_T), [p_T])$ to quantify the intrinsic correlations between the spectral shape and mean $p_T$, as well as the relative fluctuations of the particle spectrum, $\delta n(p_T)/n(p_T)$.
Finally, motivated by the weak dependence of $v_0(p_T)$ on the $\eta$-gap reported by Ref.~\cite{ATLAS:2025ztg}, the longitudinal structure and decorrelation of the radial flow is also explored .
A summary and outlook will be presented in Sec.~\ref{sec4}.

\section{The multi-stage hybrid framework}\label{sec2}
In this work, we employ the event-by-event (3+1)-dimensional CLVisc viscous hydrodynamic framework~\cite{Pang:2012he, Pang:2018zzo, Wu:2021fjf} to simulate the space-time evolution of the bulk medium and hadron production in Au+Au and U+U collisions at $\sqrt{s_{NN}}=200$ GeV. This hybrid approach integrates fluctuating initial conditions from \trento-3D~\cite{Moreland:2014oya,Soeder:2023vdn,Ke:2016jrd} with second-order dissipative hydrodynamics and a Cooper-Frye particlization routine. In the following, we briefly outline the key module of the framework.

\subsection{\trento-3D}
We generate the initial profile of the QGP using the parametric \trento-3D model~\cite{Ke:2016jrd}. At midrapidity, the distribution follows the \trento~\cite{Moreland:2014oya,Soeder:2023vdn} ansatz, where it is determined by the reduced thickness function $T_R$, defined as the generalized mean of the thickness functions of the incoming nuclei A and B,
\begin{equation}
T_R\left(p ; T_A, T_B\right) \equiv\left(\frac{T_A^p+T_B^p}{2}\right)^{1 / p}.
\end{equation}
Here, the parameter $p$ characterizes the different modes of entropy deposition.
The thickness functions of the projectile and target nuclei are expressed as
\begin{equation}
T_{A/B}(\textbf{r}_\perp)=\sum_{i=1}^{N_{A/B}} w_i T_p\left(\textbf{r}_\perp-\textbf{r}_\perp^i \right),
\end{equation}
where $T_p$ is the thickness function of each wounded nucleon with Gaussian form, and $w_i$ is a random weight sampled from a Gamma distribution with unit mean and variance $1/k$ to account for the multiplicity fluctuations observed in p+p collisions.
To approximate the IP-Glasma initial state~\cite{JETSCAPE:2020mzn,JETSCAPE:2020shq}, we set the entropy deposition parameter to $p=0$. 
The Gaussian nucleon width is fixed at $0.59$~fm, and the Gamma fluctuation parameter is set to $k=2.0$.
The positions of nucleons are sampled from the deformed Woods-Saxon distributions defined in Eqs.~(\ref{eq:defws1}) and (\ref{eq:defws2}).
We investigate the effects of deformation by varying the quadrupole ($\beta_2$) and hexadecapole ($\beta_4$) deformation parameters with $\beta_2=0,0.14,0.20,0.28$ and $\beta_4=0, 0.093$.
The nuclear radius and surface diffuseness parameters are used: $R_{0, \rm Au}=6.38\ \mathrm{fm}, a_{\rm Au}=0.535\ \mathrm{fm}$ and $R_{0, \rm U}=6.67\ \mathrm{fm}, a_{\rm U}=0.44\ \mathrm{fm}$.

The substructure of the nucleon is incorporated by modeling each nucleon as a superposition of $N 
_q$ hotspots. The nucleon thickness function is given by
\begin{equation}
T_p(\textbf{r}_\perp)=\frac{1}{N_q} \sum_{j=1}^{N_q} t_q(\textbf{r}_\perp-\textbf{r}_\perp^j),     
\end{equation}
where each hotspot $t_q$ follows a Gaussian profile of the hotspot centered at $\textbf{r}_\perp^j$. For the parameters, the number of hotspots is set to $N_q=3$, and the Gaussian width of hotspot is taken to be 0.3~fm.

To extend the \trento\ model to the longitudinal direction, 
the initial three-dimensional entropy density $s(\textbf{r}_\perp, \eta)$ in \trento-3D is constructed in a factorized form to maintain good agreement with experimental measurements at midrapidity,
\begin{equation}
    s(\textbf{r}_\perp, \eta)|_{\tau=\tau_0}=K\cdot T_R(\textbf{r}_\perp) g(\textbf{r}_\perp, y) \frac{dy}{d\eta}.
\end{equation}
Here the normalization factor $K$ will be tuned to reproduce the final particle yield ${dN/d{\eta}}$ in the most central collisions, while the function $g(\textbf{x}, y)$ is constructed from the inverse Fourier transform of its cumulant-generating function $\tilde{g}$,
\begin{equation}
\begin{aligned}
    &g(\textbf{r}_\perp, y)=\mathcal{F}^{-1}\{{\tilde{g}}(\textbf{r}_\perp, k) \},\\
    &\log \tilde{g}=i\mu k - \frac{1}{2} \sigma^2 k^2 - \frac{1}{6}i \gamma \sigma^3 k^3 + {\cdots}
\end{aligned}
\end{equation}
The dependence of the first three cumulants, $\mu, \sigma, \gamma$, on the transverse position $\textbf{r}_\perp$ is introduced via thickness function as 
\begin{equation}
\begin{aligned}
&\mu=\frac{1}{2}\mu_0 \log (T_A/T_B), \\
&\sigma=\sigma_0, \\
&\gamma=\gamma_0 \frac{T_A-T_B}{T_A+T_B}.
\end{aligned}
\end{equation}
The corresponding coefficients $\mu_0, \sigma_0, \gamma_0$ are taken as 
$\mu_0=0, \sigma_0=2.0, \gamma_0=6.0$.
Additionally, we assume the initial proper time $\tau_0$ as 0.6~fm.  

\subsection{CLVisc}
The subsequent space-time evolution of the QGP medium is simulated using the CLVisc framework~\cite{Pang:2012he, Pang:2018zzo, Wu:2021fjf}, which numerically solves the energy-momentum conservation equations,
\begin{equation}
\partial_{\mu} T^{\mu\nu} = 0,
\end{equation}
where $T^{\mu\nu}$ is the energy-momentum tensor of the QGP meidum. 
Since the QGP medium is not fully in local thermal equilibrium, 
The dissipative currents are evolved according to the Israel-Stewart-like equation of motion,
equation~\cite{Denicol:2018wdp}, 
\begin{equation}
\begin{gathered}
\tau_{\Pi} D \Pi+\Pi=-\zeta \theta-\delta_{\Pi \Pi} \Pi \theta+\lambda_{\Pi \pi} \pi^{\mu \nu} \sigma_{\mu \nu}, \\
\tau_\pi \Delta_{\alpha \beta}^{\mu \nu} D \pi^{\alpha \beta}+\pi^{\mu \nu}=\eta_v \sigma^{\mu \nu}-\delta_{\pi \pi} \pi^{\mu \nu} \theta-\tau_{\pi \pi} \pi^{\lambda\langle\mu} \sigma_\lambda^{\nu\rangle}\\+\varphi_1 \pi_\alpha^{\langle\mu} \pi^{\nu\rangle \alpha}.
\end{gathered}
\end{equation}
Here $D$ is the covariant derivative $u^\mu \partial_\mu$, $\Delta^{\mu\nu}$ is the spatial projector $g^{\mu\nu}-u^\mu u^\nu$, $\nabla^\mu=\Delta^{\mu\nu}\partial_\nu$, $\theta=\partial_{\mu}u^{\mu}$ is the expansion rate and $\sigma^{\mu\nu}$ is the symmetric shear tensor $2\nabla^{<\mu} u^{\nu>}$. 
The notation $A^{\langle\mu \nu\rangle} \equiv \Delta_{\alpha \beta}^{\mu \nu}A^{\alpha \beta}$ denotes the symmetric and traceless projection transverse to the flow velocity $u^\mu$. The corresponding projection operator is defined as $\Delta_{\alpha \beta}^{\mu \nu}=\frac{1}{2}\left(\Delta_\alpha^\mu \Delta_\beta^\nu+\Delta_\alpha^\nu \Delta^\mu{ }_\beta\right)-\frac{1}{3} \Delta^{\mu \nu} \Delta_{\alpha \beta}$. 
$\eta_\nu$ and $\zeta$ are the transport coefficients for the evolution of shear tensor and bulk pressure. $\tau_{\pi}$ and $\tau_\Pi$  are relaxation times of shear tensor and bulk pressure. $\delta_{\Pi \Pi},\lambda_{\Pi \pi},\delta_{\pi \pi},\tau_{\pi \pi}$ and $\varphi_1$ are second-order transport coefficients. The values of relaxation time and transport coefficients are taken from Ref.~\cite{Denicol:2014vaa}.
In this work, the temperature dependence of specific shear viscosity $\eta_v/s(T)$ and bulk viscosity $\zeta/s(T)$ are taken from Ref.~\cite{Bernhard:2016tnd}.
The equation of state is taken from the HotQCD Collaboration~\cite{HotQCD:2014kol}.

When the local temperature of QGP drops down to the freeze-out temperature ($T_{\rm frz}$=0.15 GeV), we use the Cornelius algorithm~\cite{Huovinen:2012is} to extract the local thermal information of hypersurface. The momentum of the thermal hadrons is sampled from the Cooper-Frye formula, 
\begin{equation}
\frac{d N_i}{d Y p_T d p_T d \phi}=\frac{g_i}{(2 \pi)^3} \int p^\mu d \Sigma_\mu f_{\mathrm{eq}}(1+\delta f),
\end{equation}
where $f_{\rm eq}$ is the thermal distribution of fermion or boson while the viscous corrections $\delta f$ to the local equilibrium distribution function follow the 14-moments method as presented in Refs.~\cite{JETSCAPE:2020mzn,McNelis:2019auj}.
In this work, we oversampled 1000 times for each hydrodynamic event.
After the thermal hadrons are produced, they undergo resonance decay procedure directly.
We run 10000 minimum bias events and perform the centrality selection based on final multiplicity of charged hadrons within $|\eta|<2.5$.

\section{Results and discussions}\label{sec3}
In this section, we present the numerical results based on the CLVisc framework. 
First, we calibrate the model by comparing its results with experimental measurements of $dN/d\eta$, $p_T$ spectra, and the mean transverse momentum $\langle [p_T] \rangle$  as functions of centrality.
Then the effects of nuclear structure on radial flow observables was explored for Au+Au and U+U collisions in various centrality classes, including the integrated radial flow $v_0$, the differential radial flow $v_0(p_T)$, the Pearson correlation coefficient ($\rho(n(p_T), [p_T])$), the relative fluctuation of $p_T$ spectrum $\frac{\delta n(p_T)}{n(p_T)}$, and the longitudinal structure of the radial flow.

\subsection{Calibration}\label{sec-calibration}
\begin{figure}
    \centering
    \includegraphics[width=\linewidth]{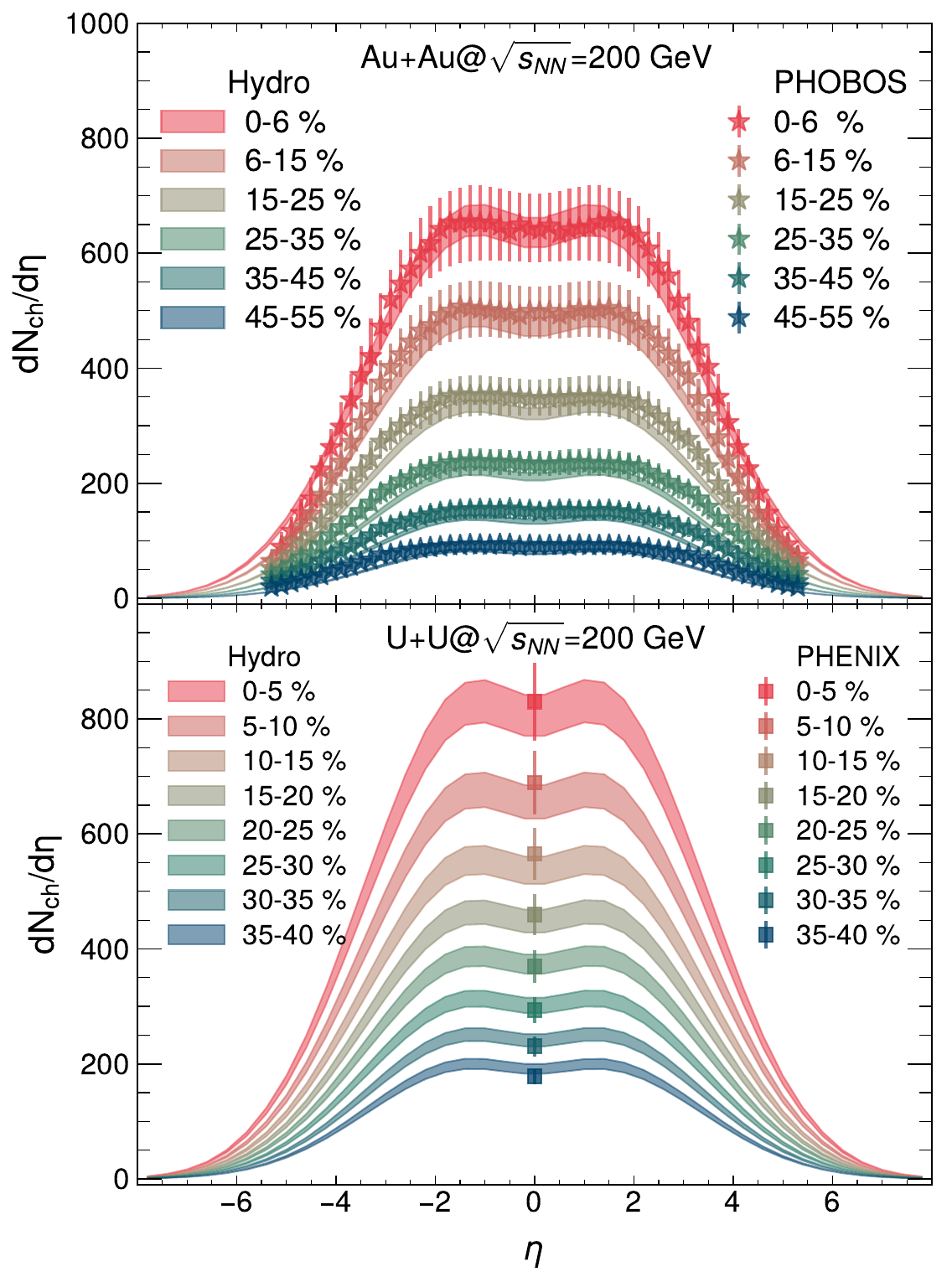}\\
    \caption{Comparison of the numerical results for the pseudorapidity dependence of charged hadrons production in various centrality classes of Au+Au and U+U collisions at $\sqrt{s_{NN}}=$200 GeV with experimental measurements from PHOBOS~\cite{PHOBOS:2010eyu} and PHENIX~\cite{PHENIX:2015tbb}.}
    \label{fig:calibration1}
\end{figure}

In Fig.~\ref{fig:calibration1}, we compare the pseudorapidity dependence of charged hadrons production in various centrality classes of Au+Au and U+U collisions at $\sqrt{s_{NN}}=$200 GeV with experimental measurements from PHOBOS~\cite{PHOBOS:2010eyu} and PHENIX~\cite{PHENIX:2015tbb}.
We can clearly see that the theoretical calculations describe experimental results pretty well in different centrality classes, not only for Au+Au collisions but also for U+U collisions, even though no additional tuning is performed for the latter\footnote{We only tuned the normalization factor $K$ to match the charged particles multiplicity in the most central (0-5\%) Au+Au collisions. The same value is used for both Au+Au and U+U collisions.}.
It should be noted that only the results for U+U collisions without deformation are shown here, because the effects of deformation on the charged hadrons multiplicity are negligible.

\begin{figure}
    \centering
    \makebox[\linewidth][l]{
    \hspace*{-2mm}
    \includegraphics[width=\linewidth]{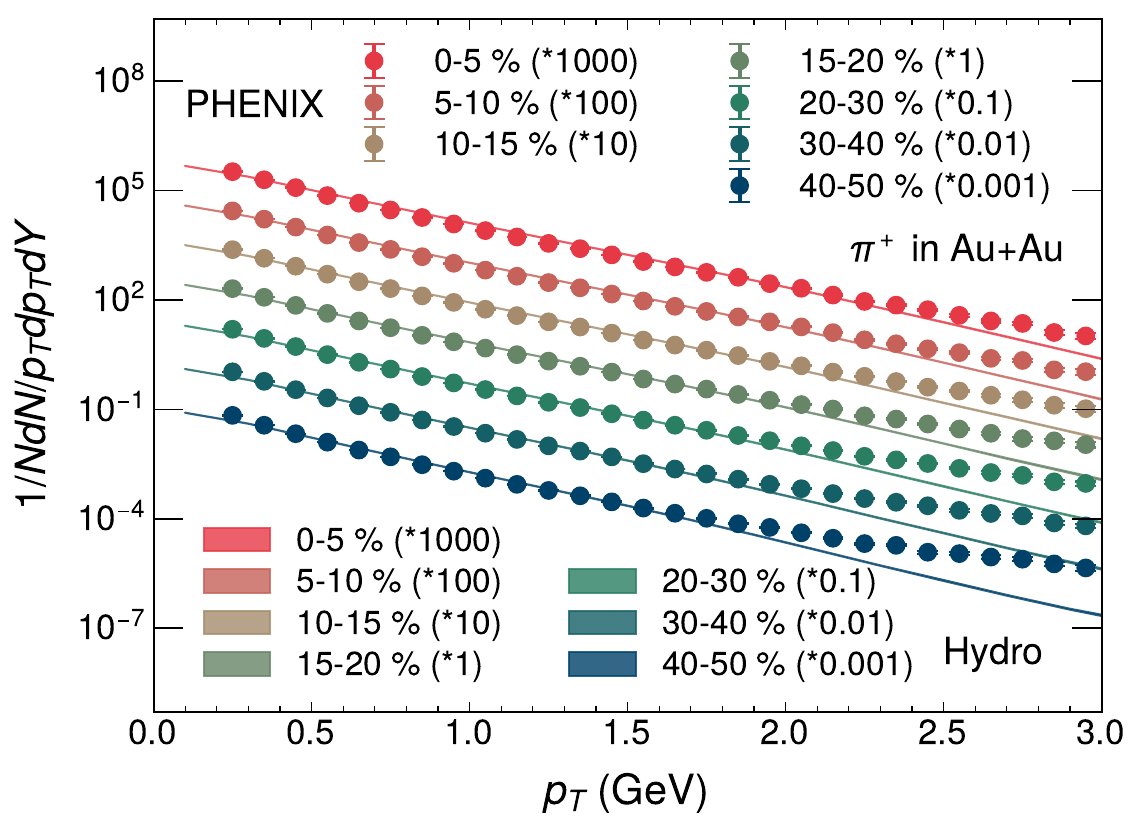}}\\
    \includegraphics[width=\linewidth]{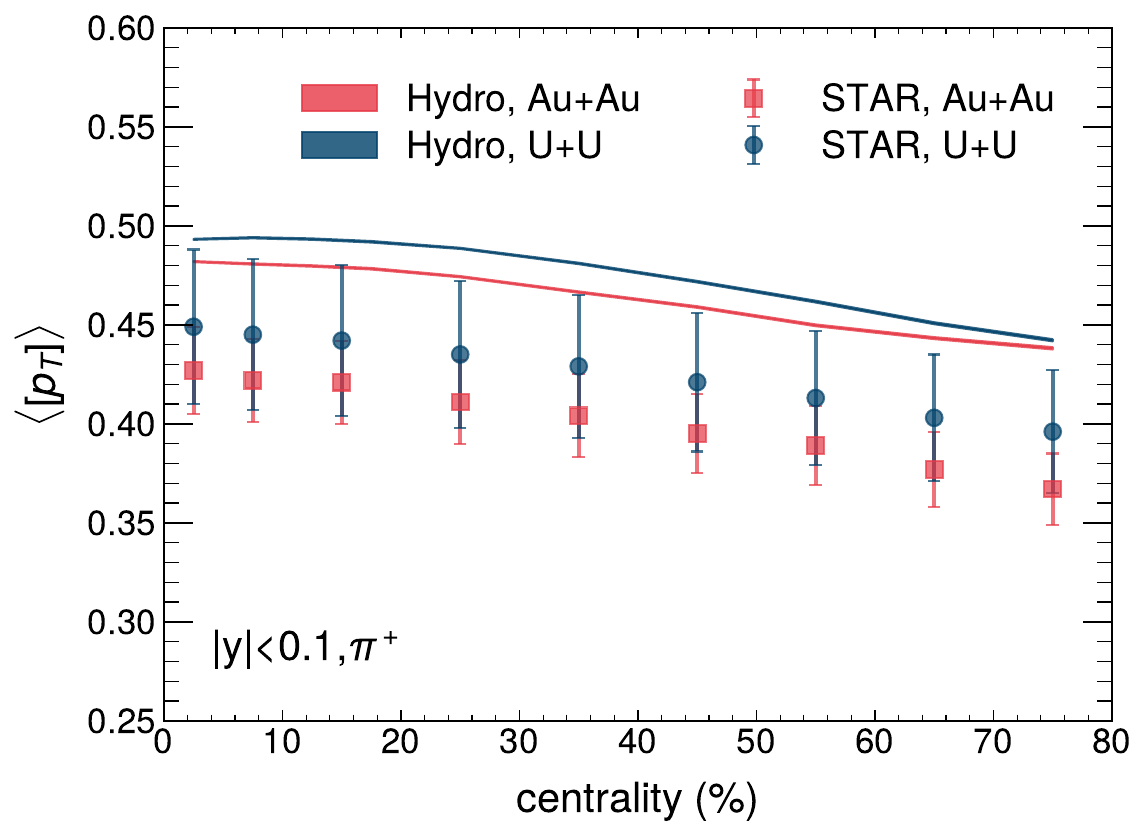}
    \caption{Comparison of numerical simulations for the $p_T$ spectra (upper) and the mean $p_T$ (lower) in various centrality classes of Au+Au and U+U collisions at $\sqrt{s_{NN}}$=200 GeV with experimental measurements from PHENIX~\cite{PHENIX:2003iij} and STAR~\cite{STAR:2008med,STAR:2022nvh}.}
    \label{fig:calibration2}
\end{figure}

\begin{figure*}
    \centering
    \includegraphics[width=\linewidth]{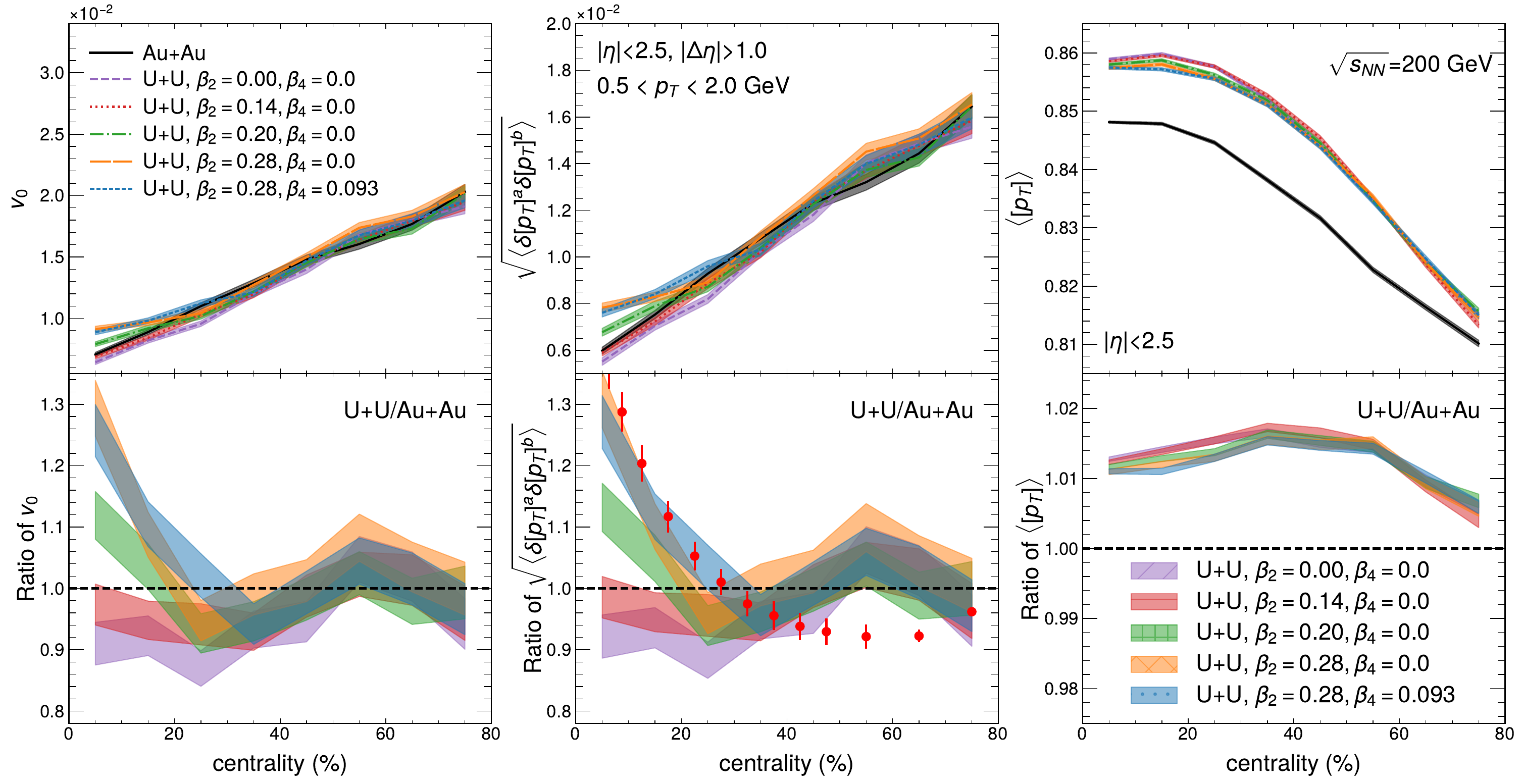}
    \caption{Centrality dependence of the integrated radial flow ($v_0$), the fluctuation and the mean value of $[p_T]$ in Au+Au and U+U collisions, and their ratios between U+U and Au+Au collisions.    Different deformation shapes of uranium are also explored. The experimental data are from~\cite{STAR:2024wgy}.
    }
    \label{fig:v0-int}
\end{figure*}

Fig.~\ref{fig:calibration2} shows the comparison of the transverse momentum spectra and the mean $p_T$ between theoretical calculations and experimental results.
Again, only results for U+U collisions without deformation are shown, since including nuclear deformation changes the mean transverse momentum by less than 1\% in our setup.
Within the error bars, the transverse-momentum spectra at low $p_T$ ($p_T < 2$ GeV) are well reproduced. The overall centrality dependence of the mean $p_T$ is also reasonably described. The current results, however, slightly underestimate the $p_T$ spectra in the range $2 < p_T < 3$ GeV and overestimate the mean $p_T$ to some extent. 
Nevertheless, this level of agreement is sufficient for a qualitative comparison of radial-flow observables.

\subsection{Radial flow observables  }
\begin{figure*}
    \centering
    \includegraphics[width=\linewidth]{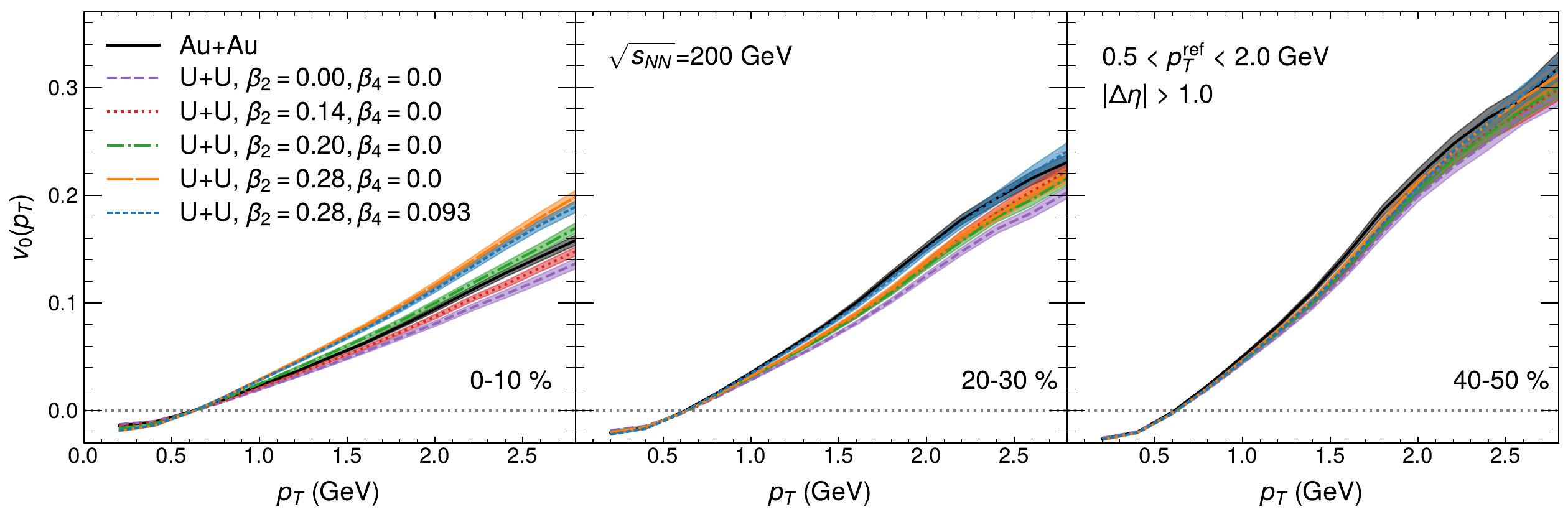}
    \caption{Transverse momentum dependence of radial flow in Au+Au and U+U collisions for several deformation parameters in three centrality bins: 0-10\%, 20-30\% and 40-50\%. }
    \label{fig:v0-pT}
\end{figure*}

Following Refs.~\cite{ATLAS:2025ztg,Schenke:2020uqq}, the $p_T$-differential and $p_T$-integrated radial-flow observables are constructed as follows.

\begin{equation}
\begin{aligned}
    &v_0(p_T)=\frac{\langle \delta n(p_T) \delta [p_T]\rangle}{\langle n(p_T)\rangle \langle [p_T]\rangle} \times {1 \over v_{0}},\\
    &v_{0}
    \equiv \frac{\sqrt{\langle (\delta [p_T])^2\rangle}}{\langle [p_T]\rangle} = \frac{\sigma_{[p_T]}}{\langle [p_T]\rangle}
    ,
\end{aligned}
\label{eq:v0}
\end{equation}
$n(p_T)$ and $[p_T]$ denote the normalized transverse-momentum spectrum and the mean transverse momentum for a given event, respectively. The symbol $\langle \cdots \rangle$ represents an average over events.  $\delta x = x - \langle x \rangle$ quantifies the deviation of $x$ from its ensemble average. $\sigma_{[p_T]}$ denotes the standard deviation of the event-by-event mean transverse momentum [$p_T$] distribution. In practice, $\delta n(p_T)$ and $\delta [p_T]$ are evaluated in two different rapidity regions, labeled “a” and “b”, which are separated by a rapidity gap ($\Delta\eta$) to suppress non-flow effects~\cite{Zhou:2015iba}.

\subsubsection{Integrated radial flow observable $v_0$}
In the left panel of Fig.~\ref{fig:v0-int}, we present the centrality dependence of the integrated radial flow observable $v_0$ and the ratio of $v_0$ between U+U and Au+Au collisions at $\sqrt{s_{NN}}=200$ GeV, where the U+U collisions are simulated with different sets of deformation parameters. 
Since $v_0$ mainly reflects the relative fluctuation of the mean transverse momentum, its separate components, namely the mean $p_T$ (right panel) and the absolute fluctuation $\sqrt{\langle \delta[p_T]^a \delta[p_T]^b \rangle}$ (middle panel), are also shown in Fig.~\ref{fig:v0-int}.
These observables exhibit clear differences between the two collision systems, reflecting the combined effects of the larger overall system size and the intrinsic deformation of the uranium nucleus.
Generally, a larger collision system  tends to generate stronger collective expansion, leading to a higher $\langle [p_T]\rangle$ and reduced event-by-event fluctuations of the mean $p_T$~\cite{Xu:2021uar}. In U+U collisions,  the nuclear deformation introduces an additional source of mean-$p_T$ fluctuations, particularly in central events, through the varying orientations of the colliding nuclei~\cite{STAR:2024wgy}.

Consistent with this picture, our calculations, the U+U/Au+Au ratios shown in the lower panel of Fig.~\ref{fig:v0-int}, show that increasing the quadrupole deformation $\beta_2$ leads to a gradual decrease in the mean $p_T$, while simultaneously enhancing both the mean-$p_T$ fluctuations and the radial flow magnitude $v_0$.
 The $\beta_2$ dependence is most important in central collisions and remains visible up to roughly 40\% centrality.
In contrast, the hexadecapole deformation $\beta_4$ has only a minor effect on the these observables. The strong sensitivity of $v_0$ to $\beta_2$ indicates that $v_0$ provides a potential additional probe of nuclear structure in deformed systems.

Finally, using $\beta_2 \simeq 0.28$ in the simulations, we qualitatively reproduce the measured ratio of mean-$p_T$ fluctuations between U+U and Au+Au collisions up to 40\% centrality, in agreement with the value inferred in the recent STAR analysis~\cite{STAR:2024wgy}.

\subsubsection{$p_T$-differential radial flow observable $v_0(p_T)$}
\begin{figure*}
    \centering
    \includegraphics[width=.85\linewidth]{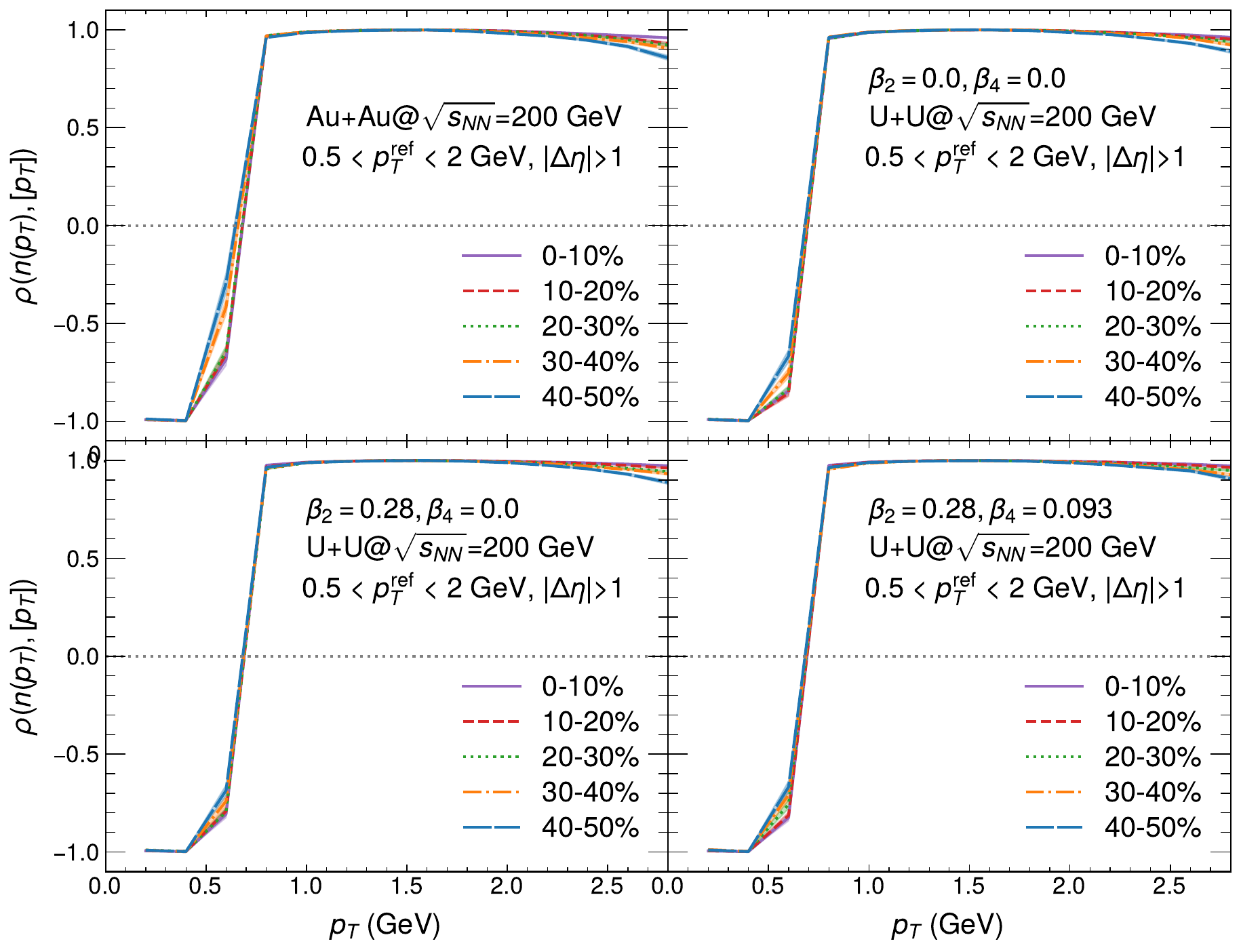}
    \caption{Pearson correlation coefficient between $n(p_T)$ and the mean $p_T$ in five centrality classes for different collision systems and deformation parameters.}
    \label{fig:rho-pT}
\end{figure*}

\begin{figure*}
    \centering
    \includegraphics[width=\linewidth]{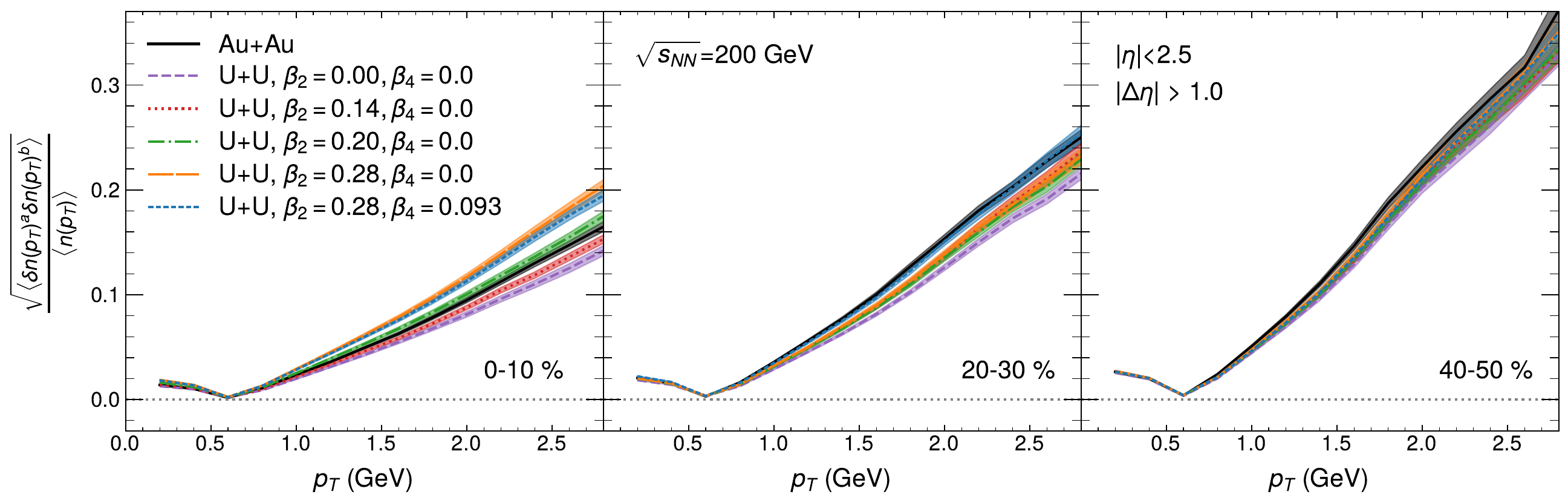}
    \caption{Transverse momentum dependence of relative fluctuation of the normalized $p_T$ spectrum, i.e. ${\sqrt{\langle \delta n(p_T)^a \delta n(p_T)^b \rangle}}/{\langle n(p_T)\rangle}$, in three centrality bins 0-10\%, 20-30\% and 40-50\% for different collision systems and deformation parameters.}
    \label{fig:n-pT}
\end{figure*}

Although the integrated radial-flow fluctuation observable $v_0$ already provides valuable constraints, the $p_T$-differential fluctuation $v_0(p_T)$ offers additional information on the microscopic properties of the QGP.
Fig~\ref{fig:v0-pT} shows the $p_T$-differential radial-flow fluctuation $v_0(p_T)$ for Au+Au and U+U collisions at $\sqrt{s_{NN}}=200$ GeV in three centrality classes (0–10\%, 20–30\%, and 40–50\%), including several choices of the uranium deformation parameters $(\beta_2,\beta_4)$. For all systems and centralities, $v_0(p_T)$ exhibits a similar shape: it increases approximately linearly with $p_T$, starting from small negative values at low $p_T$, crossing zero, and becoming positive at higher $p_T$. This behavior reflects the hydrodynamic radial expansion, where stronger radial flow boosts more particles to larger $p_T$, directly increasing the event mean $p_T$. Consequently, this leads to a positive correlation between the mean-$p_T$ fluctuations and the variations in the particle spectrum. 
 The overall magnitude of $v_0(p_T)$ grows towards more peripheral collisions, consistent with the smaller system size and larger event-by-event fluctuations in those events. A similar pattern has been observed at LHC energies~\cite{ATLAS:2025ztg, ALICE:2025iud} , but the magnitude is larger at RHIC. The larger signal at RHIC arises from stronger fluctuations at lower collision energies, in agreement with findings from anisotropic flow measurements~\cite{Giacalone:2019vwh}.

Regarding the system size and deformation dependence, $v_0(p_T)$ follows the same qualitative trends as the $p_T$-integrated $v_0$. In the absence of deformation ($\beta_2=\beta_4=0$), Au+Au collisions exhibit a larger $v_0(p_T)$ than U+U, as expected for a smaller system.
In central U+U collisions and at high $p_T$, the magnitude of $v_0(p_T)$ increases monotonically with $\beta_2$, indicating that nuclear deformation introduces additional initial-state collision modes, such as tip–tip collision, with locally higher energy density and smaller transverse size. It will enhance the correlation between the particle yield and its fluctuations at high $p_T$. Also, $\beta_4$ has only a minor impact. This sensitivity to $\beta_2$ highlights again that the radial-flow fluctuation $v_0(p_T)$ may serve as a valuable probe of nuclear structure in deformed systems.

According to previous studies, it has been known that $v_0(p_T) $
reflects fluctuations in the radial flow. To gain deeper insight into the physical origin of $v_0(p_T) $, 
we decompose $v_0(p_T)$ into two parts as follows: 
\begin{equation}
    v_0(p_T)=\rho \left( n(p_T), [p_T] \right)
    \frac{\sqrt{\langle \delta n(p_T) ^2\rangle}}{\langle n(p_T)\rangle}.
\end{equation}
Here $\rho \left( n(p_T), [p_T] \right)= \frac{\langle \delta n(p_T)\delta[p_T] \rangle}{\sqrt{\langle (\delta n(p_T) )^2\rangle \langle (\delta [p_T] )^2\rangle} }$ defines the Pearson correlation coefficient between normalized spectrum $n(p_T)$ and mean $p_T$, while the remaining factor corresponds to the relative fluctuations of $n(p_T)$. In the subsequent analysis, we will systematically examine these two components of the $v_0(p_T) $. 

In Fig.~\ref{fig:rho-pT}, we present the Pearson correlation coefficient $\rho(n(p_T), [p_T])$ for centrality classes ranging from 0–10\% to 40–50\%, shown in each panel. The top-left panel corresponds to the Au+Au collision system, while the remaining panels display results for U+U collisions at different deformation parameters. Overall, the $p_T$-dependence of $\rho(n(p_T), [p_T])$ exhibits an approximate step function: it remains near $-1$ at low $p_T$ and rises sharply toward $+1$ at high $p_T$. This behavior is a characteristic signature of collective flow. 
It reflects a strong correlation between the $[p_T]$ and $n(p_T)$, where flatter spectra are associated with larger $\langle p_T\rangle$, while steeper spectra correspond to smaller values. 
Moreover, this step-like structure shows only a weak sensitivity to centrality, collision system, and nuclear structure. 
And the zero-crossing point shows a slight centrality dependence: it shifts toward smaller $p_T$ in peripheral collisions, a trend that closely follows the centrality dependence of the mean $p_T$.
Next, we present the relative fluctuation of $n(p_T)$ in Fig.~\ref{fig:n-pT}. For a direct comparison with Fig.~\ref{fig:v0-pT}, we adopt the same analysis settings.
The relative fluctuation of $n(p_T)$ shows a deformation dependence and an overall magnitude that are similar to those of $v_0(p_T)$. Because this quantity is positive definite by definition, it remains positive even in the low-$p_T$ region where $v_0(p_T)$ becomes negative. This indicates that the main feature of the radial-flow observable $v_0(p_T)$ are largely governed by the behavior of the relative fluctuation of $n(p_T)$, which therefore encodes more fundamental properties of the QGP medium.

\subsubsection{$\eta$-differential radial flow observable $v_0(\eta)$ and longitudinal decorrelation $r_0(\eta)$}
\begin{figure}
    \centering
    \includegraphics[width=\linewidth]{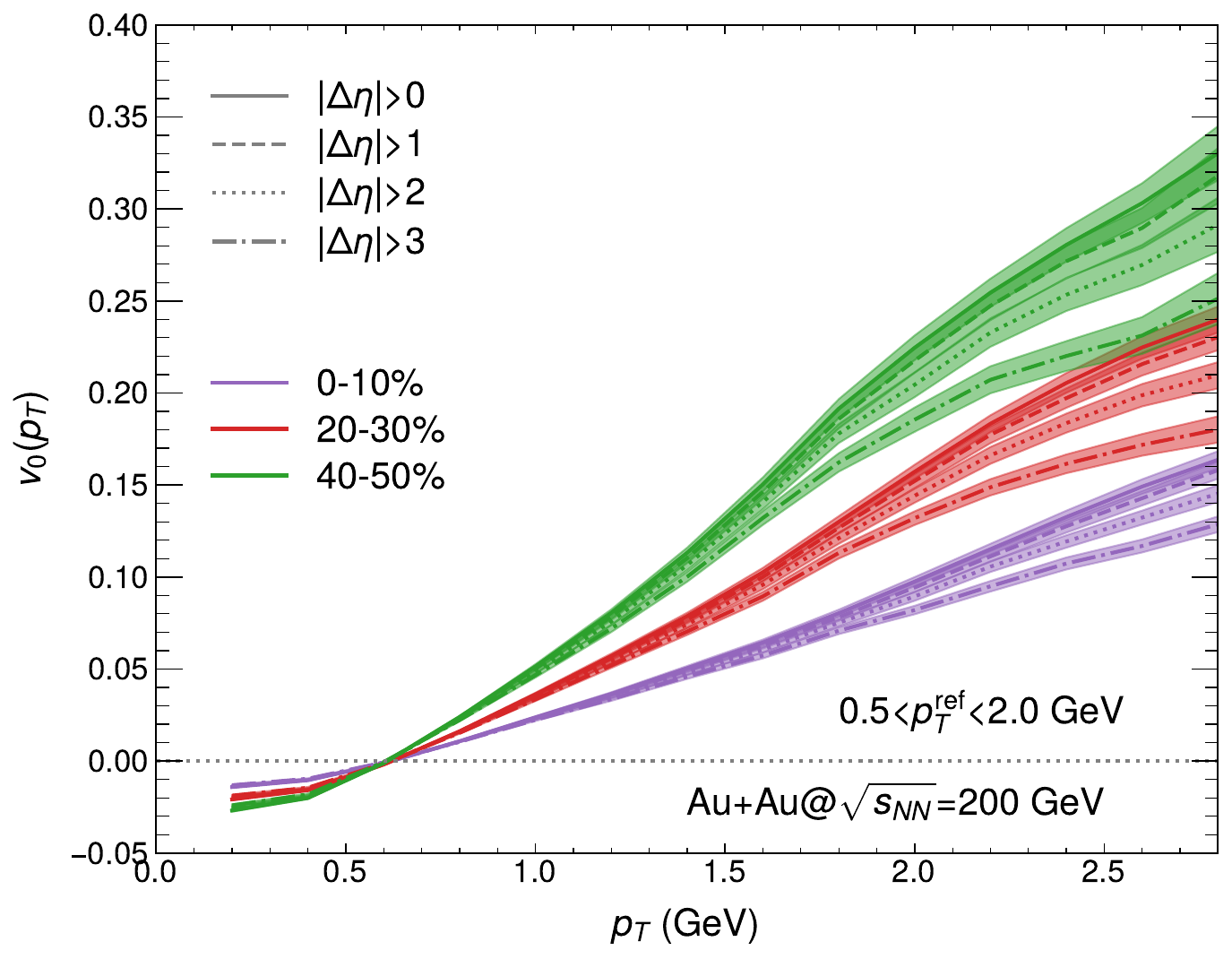}
    \caption{Impact of the rapidity gap on $v_0(p_T)$ for four centrality percentiles in Au+Au collisions at $\sqrt{s_{NN}}$=200 GeV.}
    \label{fig:v0-etagap}
\end{figure}

Although radial-flow fluctuations in the transverse plane have attracted considerable attention, it is equally important and interesting to investigate their longitudinal structure. One way to probe this longitudinal structure is to study the dependence of $v_0(p_T)$ on the pseudorapidity gap $\Delta\eta$, as proposed in Ref.~\cite{ATLAS:2025ztg}. The ATLAS data show almost no $\Delta\eta$ dependence for $v_0(p_T)$ in central collisions and only a weak dependence in peripheral events. We have examined the effect of the rapidity gap on $v_0(p_T)$ in our model simulations, focusing on Au+Au collisions at $\sqrt{s_{NN}}=200$ GeV, as shown in Fig.~\ref{fig:v0-etagap}. Surprisingly, a clear separation between different $\eta$-gaps is observed for all centrality, with a trend opposite to that reported by ATLAS: larger $\eta$-gaps suppress the magnitude of $v_0(p_T)$ at high $p_T$, in analogy to the behavior known from anisotropic-flow measurements~\cite{ALICE:2017xzf}. In experimental data such a $\Delta\eta$ dependence is often attributed to non-flow contributions, whereas in hydrodynamic simulations non-flow effects are expected to be much smaller. A possible explanation is that longitudinal fluctuations of radial flow are intrinsically stronger at RHIC energies than at the LHC.

\begin{figure*}
    \centering
    \makebox[\linewidth][l]{\hspace*{4mm}
    \includegraphics[width=.96\linewidth]{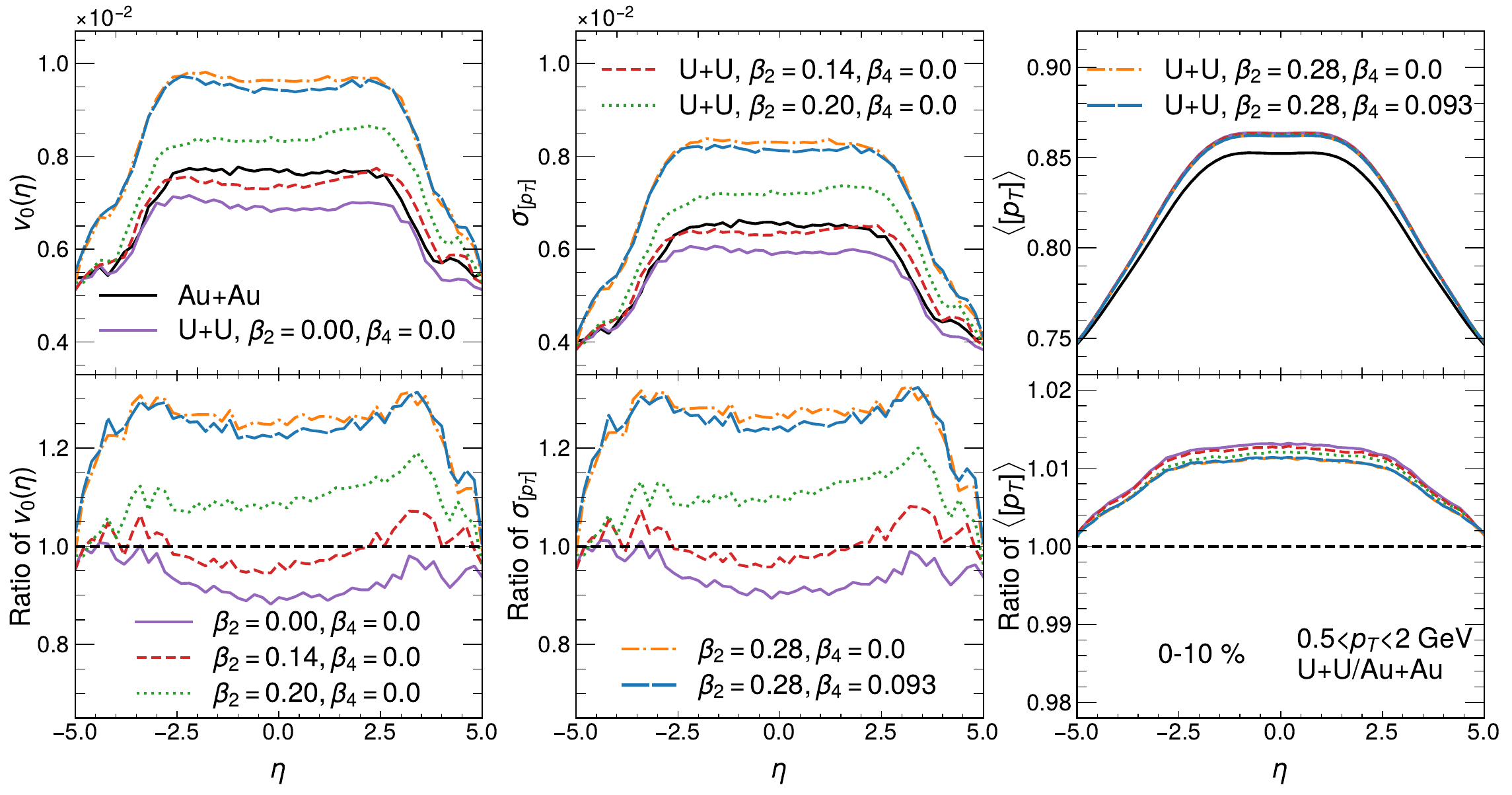}}\\
    \caption{$\eta$ dependence of $v_0$, $\sigma_{[p_T]}$ and $\langle [p_T] \rangle$, as well as respective ratio in Au+Au and U+U central collisions. Different deformation shapes of uranium are also explored.}
    \label{fig:pT-eta}
\end{figure*}
An alternative approach to probing the longitudinal structure is to define the $\eta$-dependent radial flow observable, $v_0(\eta) = \frac{\sigma_{[p_T]}} {\langle [p_T] \rangle}(\eta)$, in analogy to Eq.~(\ref{eq:v0}). Figure.~\ref{fig:pT-eta} presents the $\eta$ dependence of the radial flow $v_0(\eta)$, the absolute fluctuation $\sigma_{[p_T]}$ and the mean transverse momentum $\langle [p_T]\rangle$, together with their ratios between Au+Au and U+U collisions in the 0–10\% centrality class.
In analogy with rapidity-dependent anisotropic flow,  radial flow $v_0(\eta)$, the mean $p_T$ and its fluctuations $\sigma_{[p_T]}$ all exhibit a broad plateau around midrapidity and decrease toward forward and backward rapidities. At the same time, in the central rapidity region the effects of the collision system size and nuclear structure show a dependence very similar to that observed at midrapidity.

Finally, inspired by the longitudinal decorrelation observables for anisotropic flow, we define a similar longitudinal decorrelation observables for radial flow, denoted $r_0(\eta)$.
\begin{equation}\label{def:r0}
r_0(\eta)\equiv 
\frac{\langle \delta [p_T]_{-\eta}\cdot \delta[p_T]_{\eta^{\rm ref}} \rangle}{ \langle \delta [p_T]_{\eta}\cdot \delta [p_T]_{\eta^{\rm ref}} \rangle}
\approx \frac{\rho(-\eta, \eta^{\rm ref})}{\rho(\eta, \eta^{\rm ref})}.
\end{equation}
In the last step, we assume that the fluctuation of the mean $p_T$ is symmetric along the longitudinal direction, i.e. $\sigma_{[p_T], \eta}=\sigma_{[p_T], -\eta}$, which has already been confirmed in Fig.~\ref{fig:pT-eta}. 
In the following calculations, the reference rapidity is chosen as $\eta^{\rm ref} \in [3,5]$. 
According to the definition, any deviation of $r_0(\eta)$ from unity indicates a decorrelation of $\delta [p_T]$ between $\eta$ and $-\eta$, and $r_0$ should always be smaller than one.

\begin{figure*}
    \centering
    \includegraphics[width=\linewidth]{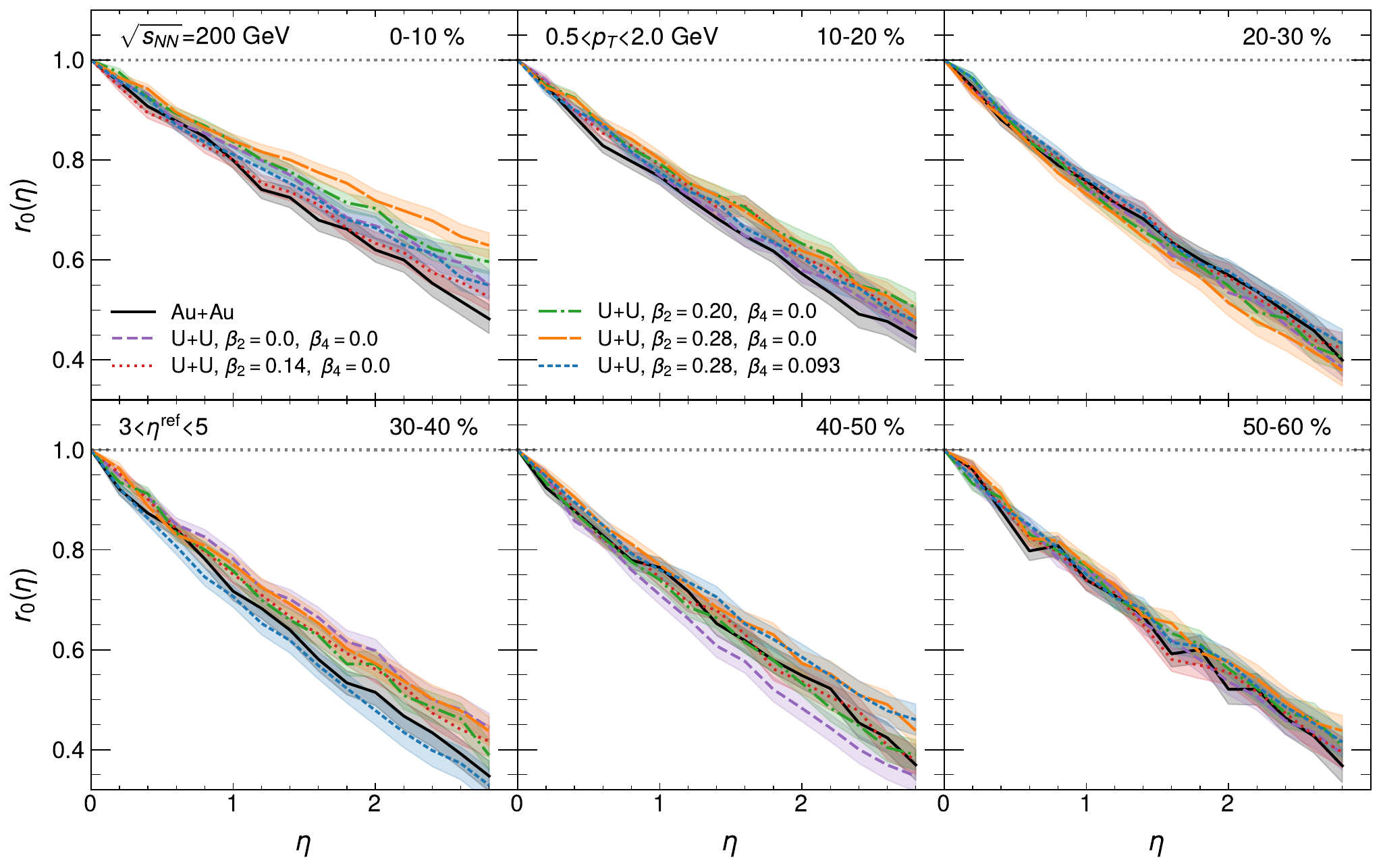}
    \caption{Longitudinal decorrelation, $r_0(\eta)$, in various centrality classes of Au+Au and U+U collisions, where various deformations of uranium are considered.}
    \label{fig:r0}
\end{figure*}

In Fig.~\ref{fig:r0}, the longitudinal decorrelation observable $r_0(\eta)$ is shown for six centrality classes in Au+Au and U+U collisions, including several choices of the uranium deformation. Around midrapidity, $r_0(\eta)$ exhibits an approximately linear dependence on $\eta$. Its centrality dependence is qualitatively different from that of the longitudinal decorrelation of elliptic flow: since radial flow is only weakly affected by the initial collision geometry, the behavior of $r_0(\eta)$ is more similar to that of triangular flow, with weaker decorrelation (smaller slope) in the most central collisions where the system size is largest. In addition, we find that large $\beta_2$ suppresses the longitudinal decorrelation of radial flow, whereas $\beta_4$ enhances it in central collisions. Furthermore, a smaller system size ($R_{\rm Au} < R_{\rm U}$) tends to result in stronger longitudinal decorrelation for the non-deformed case. Therefore, this new observable provides an additional constraint on the longitudinal structure of the QGP.

\section{Summary and outlook}\label{sec4}
In this work, we have presented a study of $p_T$ integrated and differential radial flow, as well as its longitudinal structure, based on the (3+1)-dimensional CLVisc hydrodynamics framework with the \trento-3D initial conditions. In particular, we have examined the effect of nuclear deformation on various radial flow observables.

Firstly, our calculations can reproduce the experimental results for the pseudorapidity dependence of particle production across various centrality classes in both Au+Au and U+U collisions. 
Although the model slightly overestimates the mean $p_T$, it captures its centrality dependence.
We find that the effects of nuclear deformation on both the multiplicity and the mean $p_T$ are negligible.

Regarding the impact of nuclear structure on radial flow, 
we observe that quadrupole deformation $\beta_2$ tends to enhance both the integrated radial flow $v_0$ and the $p_T$ differential radial flow $v_0(p_T)$ in central and mid-central collisions. This enhancement can be explained by the random orientations of the deformed nuclei.
We also find that the intrinsic correlation between the transverse momentum spectrum and the mean $p_T$, quantified by the Pearson correlation coefficient $\rho(n(p_T), [p_T])$, exhibits a simple step-like pattern across all centrality classes and collision systems and shows a little sensitivity to nuclear structure.
Furthermore, the exploration of the longitudinal structure of radial flow reveals that $v_0(\eta)$, along with its two components, the mean $p_T$ and $\sigma_{[p_T]}$, decreases toward the forward and backward rapidity regions.
The effects of the collision system size and nuclear structure in the large rapidity regions are similar to those at mid-rapidity, whereas the longitudinal decorrelation of radial flow, $r_0(\eta)$, exhibits a more interesting behavior. 
In central collisions, large $\beta_2$ tends to suppresses the decorrelation of radial flow, while hexadecapole deformation $\beta_4$ and the smaller system size of Au enhance the decorrelation effect. Moreover, the decorrelation effects become stronger toward more peripheral collisions.

Overall, our results indicate that $v_0$ is a valuable potential probe for constraining initial nuclear geometry. Looking forward, extending this study to asymmetric collision systems, such as Pb+Ne and Pb+O~\cite{Giacalone:2024luz}, will further deepen our understanding of the nuclear structure of light nuclei from the perspective of high-energy nuclear physics.

\section{ACKNOWLEDGMENTS}

J.Z. would like to thank Hao-Wei Deng and Yan-Ran Li for productive discussions. This work is supported in part by Natural Science Foundation of China (NSFC) under Grant No. 12225503. 
J.Z. is supported in part by China Scholarship Council (CSC) under Grant No. 202306770009. X.-Y.W. is supported in part by the Natural Sciences
and Engineering Research Council of Canada (NSERC)
[SAPIN-2020-00048 and SAPIN-2024-00026], in part by
US National Science Foundation (NSF) under grant number OAC-2004571.
The numerical calculations have been performed on the GPU cluster in the Nuclear Science Computing Center at Central China Normal University (NSC3).

\bibliographystyle{apsrev4-1}
\bibliography{refs}

\appendix

\addcontentsline{toc}{section}{Appendix}

\section{The sensitivity of $P([p_T])$ to system size and nuclear shape}
\begin{figure}[b]
    \centering
    \includegraphics[width=\linewidth]{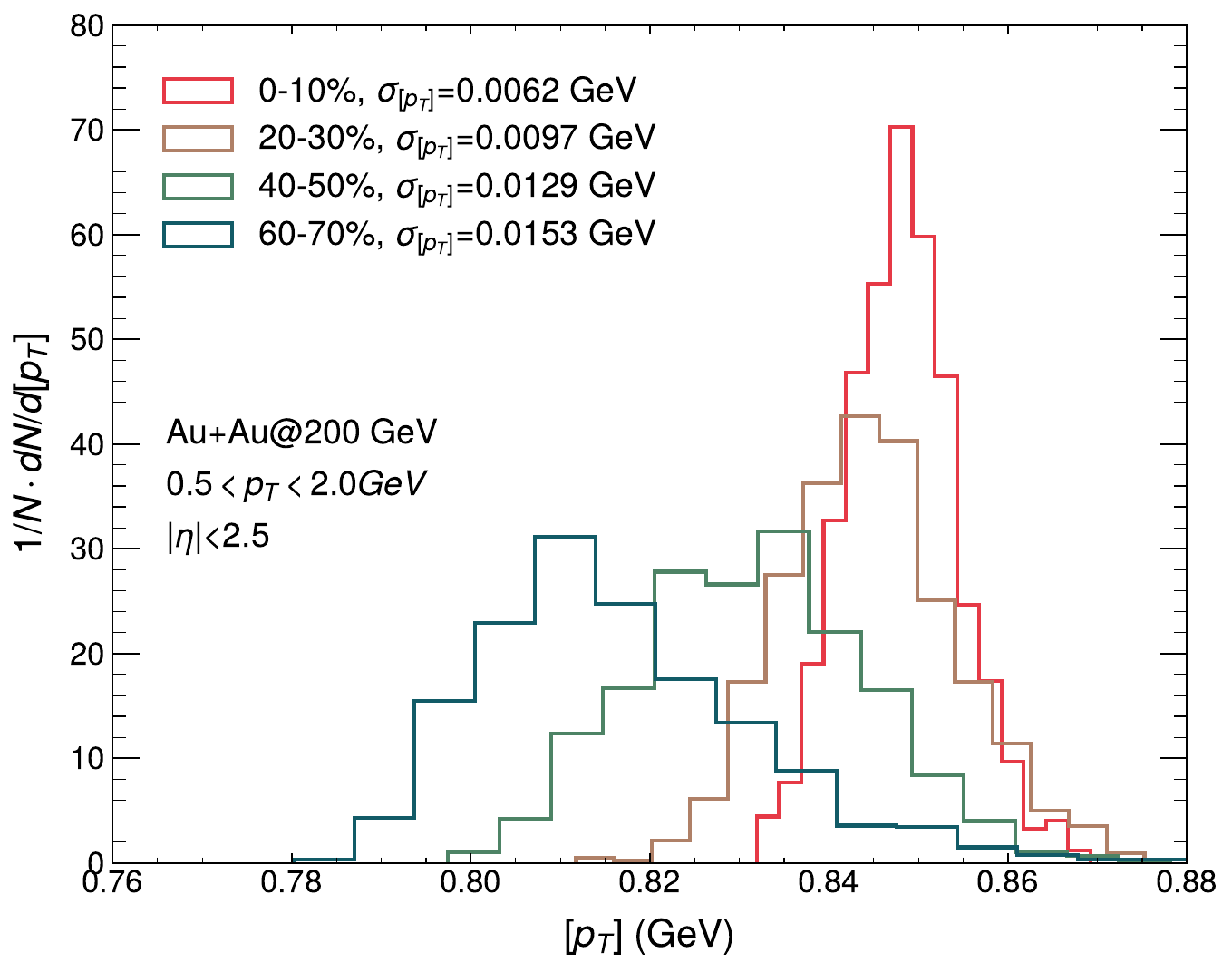}
    \includegraphics[width=\linewidth]{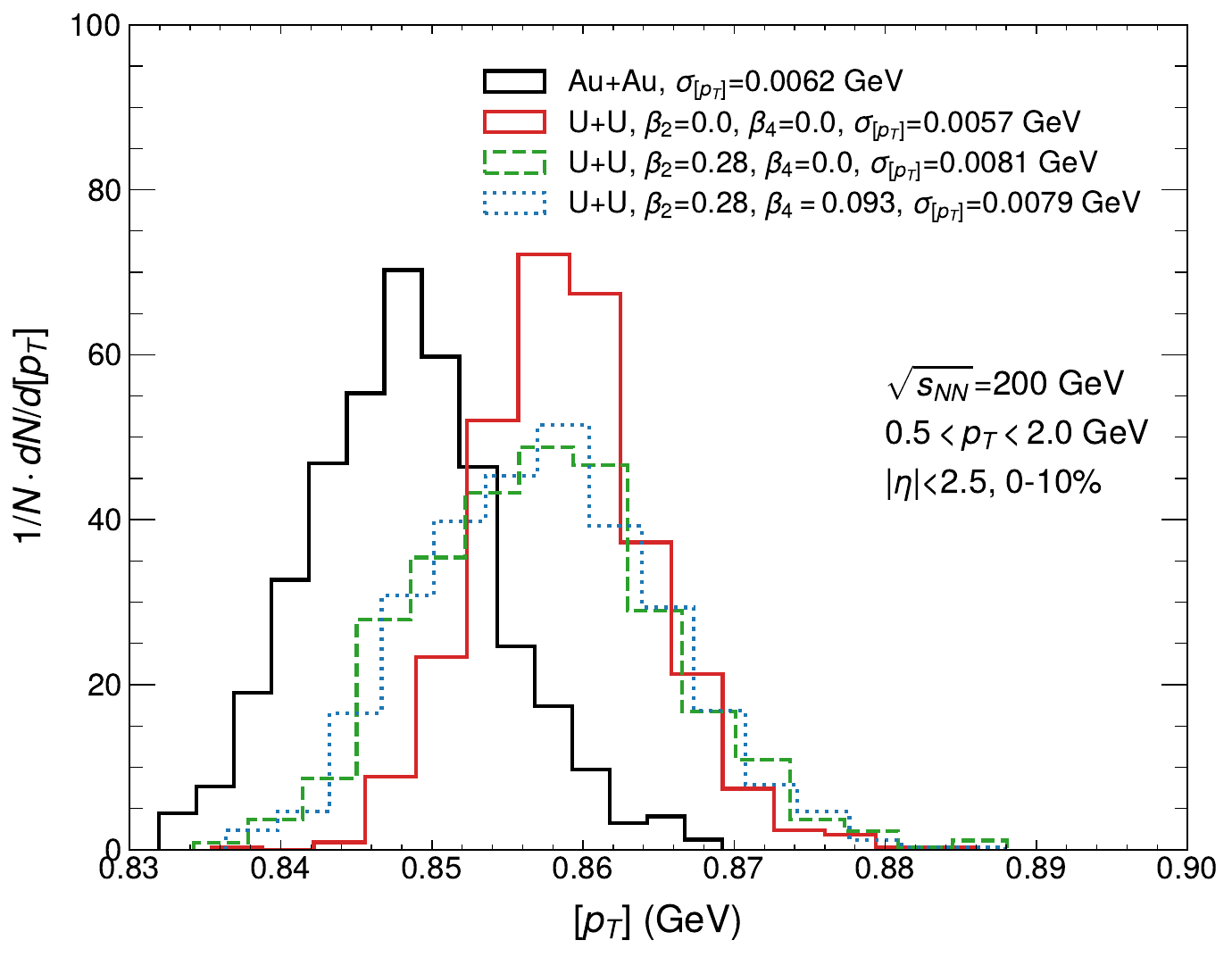}
    \caption{(\textit{Upper}): Event-by-event distributions of mean $p_T$ in four centrality classes of Au+Au collisions at $\sqrt{s_{NN}}$=200 GeV; (\textit{Lower}): Effects of nuclear structure on the event-by-event distributions of mean $p_T$ in central collisions.}
    \label{fig:meanpT-dist-1}
\end{figure}

In the upper panel of Fig.~\ref{fig:meanpT-dist-1}, we show the event-by-event distributions of the mean $p_T$ in four centrality classes of Au+Au collisions at $\sqrt{s_{NN}}$=200 GeV. It can be clearly observed that, as the collisions become more peripheral, the mean $p_T$ decreases while its fluctuation becomes larger. Furthermore, the effects of nuclear structure in central collisions are illustrated in the lower panel of Fig.~\ref{fig:meanpT-dist-1}, where the peak positions of distributions of the mean $p_T$ of U+U collisions are consistently located at higher $p_T$ than those of Au+Au collisions, owing to the slightly larger size of uranium nuclei. A closer inspection of the mean $p_T$ distributions for different deformations in U+U collisions reveals that a larger quadrupole deformation $\beta_2$ results in a broader distribution, which is consistent with the results shown in Fig.~\ref{fig:v0-int}.

\begin{figure*}[htb]
    \centering
    \includegraphics[width=0.8\linewidth]{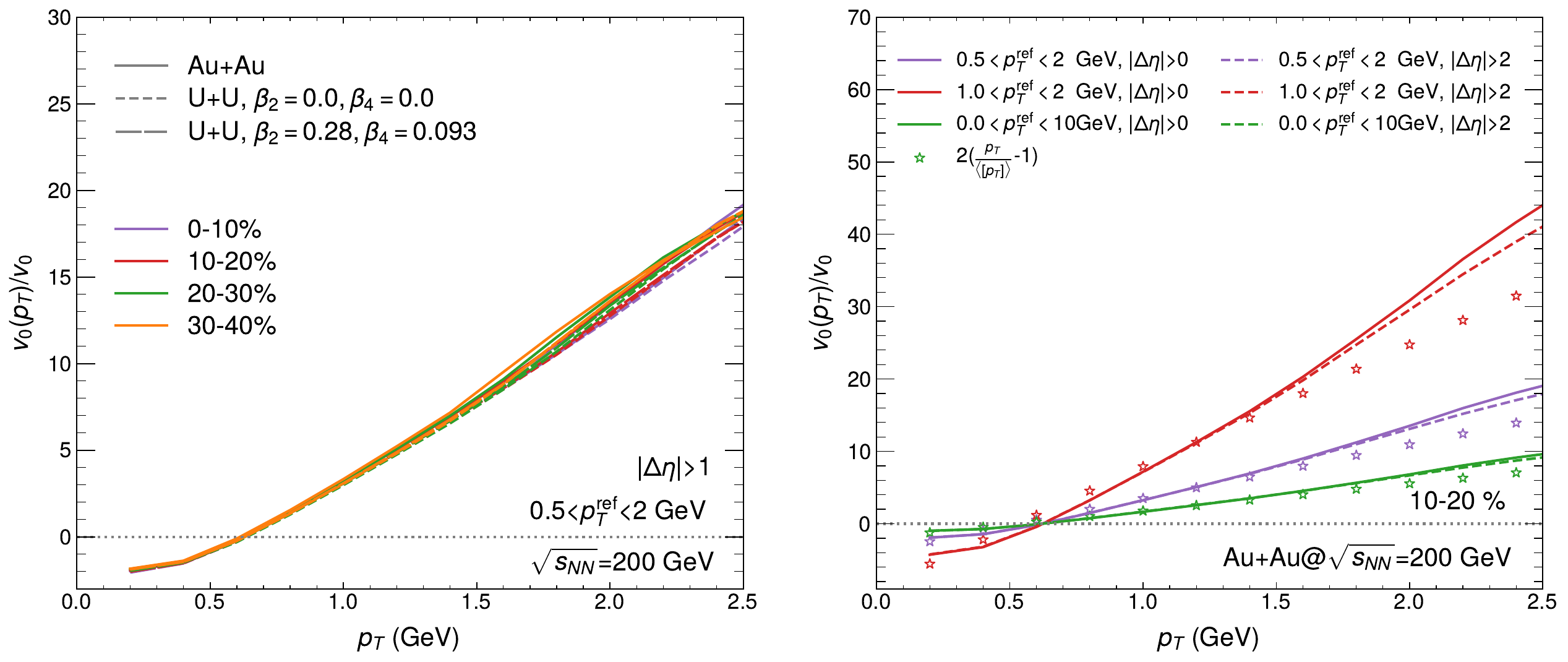}
    \caption{(\textit{Left}): Transverse momentum dependence of the scaled radial flow $v_0(p_T)/v_0$ in various centrality bins of Au+Au collisions and U+U collisions. (\textit{Right}): Comparison of the scaled radial flow $v_0(p_T)/v_0$ from theoretical simulations with the results from a simple model with different $p_T$ cut or $\eta$-gap. 
    }
    \label{fig:v0-pT-check}
\end{figure*}

\section{Scaled differential radial flow $v_0(p_T)/v_0$ }

In the left panel of Fig.~\ref{fig:v0-pT-check}, we examine the sensitivity of the scaled radial flow $v_0(p_T)/v_0$ to nuclear deformation
We found that this observable is insensitive to the deformation parameter.
Then, we analyze $v_0(p_T)/v_0$ based on a simple model in which the event-by-event fluctuations of the normalized transverse-momentum spectrum originate solely from fluctuations of the mean transverse momentum $[p_T]$~\cite{Schenke:2020uqq, Gardim:2019iah}, like,
\begin{equation}
    n(p_T;[p_T]) = 4 p_T \frac{e^{-2p_T/[p_T]}}{[p_T]^2}.
\end{equation}
Thus, the scaled radial flow can be calculated analytically as $v_0(p_T)/v_0=2(\frac{p_T}{\langle [p_T]\rangle}-1)$.
The results indicate that $v_0(p_T)/v_0$ is almost independent of centrality and nuclear structure in the low $p_T$ region ($<$ 1.5 GeV). It implies that $v_0(p_T)/v_0$ is insensitive to the details in initial state and thus is a clean probe of transport properties of the QGP~\cite{ATLAS:2025ztg}. 
Furthermore, a comparison of the $v_0(p_T)/v_0$ from this simple model~\cite{Schenke:2020uqq, Gardim:2019iah} with our calculation for 10-20\% Au+Au collisions shows good consistency in the low $p_T$ region ($< 1.5$ GeV) across different rapidity gaps and reference $p_T$ ranges.

\end{document}